\documentstyle[preprint,aps]{revtex}
\begin{document}
\draft
\vskip 2.5cm
\title{\bf Factorization of gravitational
           Compton scattering amplitude \\
           in the linearized version of general relativity}
\author{S. Y. Choi, J. S. Shim, and H. S. Song}
\address{Center for Theoretical Physics
         and Department of Physics,\\
         Seoul National University, Seoul 151-742, Korea}

\maketitle
\vskip 4cm
\begin{abstract}
Gravitational Compton scattering process with a massive fermion
is studied in the context of the linearized gravity.
Gravitational gauge invariance and graviton transversality
cause the transition amplitude to be factorized into that of
scalar QED Compton scattering and that of fermion QED Compton
scattering with an overall kinematical factor.
The factorization is shown explicitly and its physical
implications are discussed.
\end{abstract}
\vskip 0.3cm

Newton's long-ranged gravity force law  suggests the
force to be mediated by a massless particle
called a graviton and several observations of universal
gravitational attraction support  a spin-2 graviton
picture \cite{Feynman}.

Until now, there is no complete theory of quantum gravity.
One of its main problems is that Einstein gravity is
nonrenormalizable \cite{Deser}
because the gravitational constant $G_N$ has dimensions of
inverse mass square.
This aspect implies that the Einstein theory of gravitation
seems to be more like other nonrenormalizable effective
theories such as the Fermi theory of weak interaction.
Incidentally, Weinberg \cite{Weinberg} showed that it is
impossible to construct a Lorentz-invariant quantum theory
of particles of mass zero and helicity $\pm 2$ without
introducing some sort of gauge invariance into the theory.
On the other hand, even if it is nonrenormalizable,
the classical theory of gravitational radiation in the
linearized version of general relativity is well known to
have gauge invariance related to the general covariance of
the full theory. The fact uniquely determines the long
wavelength structure of quantum gravitation to be that of
Einstein's theory \cite{Weinberg,Boul}.

Several graviton interaction processes have been studied
previously. The cross section of tree-level gravitational
Compton scattering $ge\rightarrow ge$ was calculated in
Ref.~\cite{Gupta}.
The cross section of an annihilation process
$e\bar{e}\rightarrow gg$ was calculated some time ago in
Ref.~\cite{Vlad}. However, none of the two calculations were
consistent with the result of Voronov \cite{Voronov},
whose Lagrangian is gravitationally gauge invariant.
Whatever the theory is, it is so crucial to maintain
general covariance in the theory that one has to introduce
gravitational gauge invariance in the interaction
Lagrangian of linearized gravity \cite{Weinberg}.
The fact justifies the use of the same interaction Lagrangian
as in Ref.~\cite{Voronov}.

Gravitational gauge invariance and transversality of graviton
tensors cause the amplitude of gravitational Compton scattering
to be completely factorized into that of scalar QED Compton
scattering and that of fermion QED Compton scattering, both of
which are gauge invariant. One overall kinematical factor clearly
exhibits the unitarity violation of the process at quite high
energies.
In this report, we show the factorization explicitly and discuss
its physical implications

The Lagrangian ${\cal L}$ \cite{Voronov} of the interaction of
a massive fermion with the gravitational field has two terms :
the Lagrangian ${\cal L}_g$ of the pure
gravitational field and the general covariant Lagrangian
${\cal L}_f$ of the massive fermion
\begin{eqnarray}
{\cal L}&=&{\cal L}_g+{\cal L}_f,\nonumber\\
{\cal L}_g&=&\frac{1}{\kappa^2}\sqrt{-g}g^{\mu\nu}
            \left[{\Gamma_{\mu\nu}}^\alpha
            {\Gamma_{\alpha\beta}}^{\beta}
           - {\Gamma_{\mu\alpha}}^\beta
            {\Gamma_{\nu\beta}}^{\alpha}\right],\nonumber\\
{\cal L}_f&=&\sqrt{-g}\left[\frac{i}{2}(\bar{\psi}\gamma^i
             \bigtriangledown_i\psi
            -\tilde{\bigtriangledown}_i\bar{\psi}
             \gamma^i\psi)-m\bar{\psi}\psi\right].
\end{eqnarray}
where $g={\rm det}[g_{\mu\nu}]$, $\kappa=\sqrt{16\pi G_N}$,
$\gamma^i=\lambda^i_\mu \gamma^\mu$ with a certain reference
$\lambda^i_\mu$ and ordinary Dirac matrices $\gamma^\mu$, and
\begin{eqnarray}
\bigtriangledown_i\psi=\psi_{,i}-\Gamma_i\psi,\ \
\tilde{\bigtriangledown}_i\bar{\psi}=\bar{\psi}_{,i}
+\bar{\psi}\Gamma_i,
\end{eqnarray}
where $\Gamma_i=\gamma^\mu\gamma^\nu
                \lambda^k_{\mu,i}\lambda_{\nu k}$.
With general covariance preserved, it is convenient to choose
the reference uniquely connected with the metric tensor.
The choice of reference in Ref.~\cite{Ogie} is
\begin{eqnarray}
\lambda_{\mu i}=\sqrt{g_{\mu i}},
\end{eqnarray}
with the square taken in the matrix sense.

Let us consider a symmetric and traceless tensor $h_{\mu\nu}$
representing a deviation of the metric tensor $g_{\mu\nu}$ from
the flat tensor $\eta_{\mu\nu}$:
\begin{eqnarray}
g_{\mu\nu}=\eta_{\mu\nu}+\kappa h_{\mu\nu},\label{linear}
\end{eqnarray}
where  the flat tensor $\eta_{\mu\nu}$ is the Minkowski tensor
with signature ($+,-,-,-$).
Expanding the Lagrangian ${\cal L}_g$  up to the second order
of $\kappa$, we have
\begin{eqnarray}
{\cal L}_g&=&{\cal L}^0_g+\kappa{\cal L}^1_g
           +\kappa^2{\cal L}^2_g,\nonumber\\
{\cal L}^0_g&=&-\frac{1}{4}\left[2\partial_\alpha h_{\mu\nu}
             \partial^\nu h^{\mu\alpha}-\partial_\alpha h
             \partial_\mu h^{\mu\alpha}+\partial^\mu h
             \partial_\mu h-\partial_\alpha h^{\mu\nu}
             \partial^\alpha h_{\mu\nu}\right],\nonumber\\
{\cal L}^1_g&=&\frac{1}{4}\left[(h_{\mu\nu}-\frac{1}{2}
             \eta_{\mu\nu}h)\eta_{\alpha\gamma}\eta_{\beta\delta}
             +\eta_{\mu\nu}\eta_{\beta\delta}h_{\alpha\gamma}
             +\eta_{\mu\nu}\eta_{\alpha\gamma}h_{\beta\delta}\right]
             \nonumber\\
            &\times&
             \left[2\partial^\beta h^{\gamma\mu}\partial^\alpha
             h^{\delta\nu}-2\partial^\delta h^{\beta\gamma}
             \partial^\alpha h^{\mu\nu}+\partial^\nu
             h^{\alpha\gamma}\partial^\mu h^{\beta\delta}
             -\partial^\mu h^{\beta\gamma}\partial^\nu
             h^{\alpha\delta}\right],
\end{eqnarray}
and ${\cal L}^2_g $ is a certain complicated polynomial of fourth
degree in $h_{\mu\nu}$ and $h=\eta^{\mu\nu}h_{\mu\nu}$. The full
expression of ${\cal L}^2_g$ can be found in Ref.~\cite{Gross}.
For the general covariant Lagrangian of the massive fermion we have
in an analogous manner
\begin{eqnarray}
{\cal L}_f&=&{\cal L}^0_f+\kappa{\cal L}^1_f
           +\kappa^2{\cal L}^2_f,\nonumber\\
{\cal L}^0_f&=&\frac{i}{2}\left[\bar{\psi}\gamma^\mu
             \partial_\mu \psi-\partial_\mu\bar{\psi}\gamma^\mu
             \psi\right]-m\bar{\psi}\psi,\nonumber\\
{\cal L}^1_f&=&\frac{1}{2}h{\cal L}^0_f-\frac{i}{4}h_{\mu\nu}
             \left[\bar{\psi}\gamma^\nu\partial^\mu \psi-
             \partial^\mu\bar{\psi}\gamma^\nu\psi\right],\nonumber\\
{\cal L}^2_f&=&\frac{1}{8}(h^2-2h^{\mu\nu}h_{\mu\nu})
             {\cal L}^0_f+\frac{i}{16}(3h_{\mu\alpha}h^\alpha_\nu-2h
             h_{\mu\nu})(\bar{\psi}\gamma^\mu\partial^\nu\psi
             -\partial^\nu\bar{\psi}\gamma^\mu\psi)\nonumber\\
             &+&\frac{i}{32}\left[2(h_{\mu\alpha}\partial^\alpha
             h^{\mu\nu}-h^{\mu\nu}\partial^\alpha h_{\mu\alpha})
             (\bar{\psi}\gamma_\nu\psi)+
             (h^{\mu\nu}\partial^\alpha h^\beta_\mu
             -h^{\beta\mu}\partial^\alpha h^\nu_\mu)
             (\bar{\psi}\gamma_\alpha \gamma_\beta
             \gamma_\gamma\psi)\right].
\end{eqnarray}

The gravitational Compton scattering process $g e\rightarrow ge $
is made up of four Feynman diagrams (See Fig.\ \ref{Feynman Diagrams})
and the explicit form of its tree-level transition amplitude
${\cal M}_{gg}$ is determined from the Lagrangian ${\cal L}$
as
\begin{eqnarray}
{\cal M}_{gg}&=&{\cal M}_a
              +{\cal M}_b+{\cal M}_c+{\cal M}_d,\nonumber\\
{\cal M}_a&=&-\frac{\kappa^2}{4}\frac{(p_1\cdot\epsilon_1)(p_2\cdot
            \epsilon^*_2)}{(p_1\cdot k_1)}\bar{u}(p_2)
            \left[\not\!{\epsilon^*_2}(\not\!{p_1}+\not\!{k_1}+m)
            \!\!\not\!{\epsilon_1}\right]u(p_1),\nonumber\\
{\cal M}_b&=&\frac{\kappa^2}{4}\frac{(p_2\cdot\epsilon_1)(p_1\cdot
            \epsilon^*_2)}{(p_1\cdot k_2)}\bar{u}(p_2)
            \left[\not\!{\epsilon_1}(\not\!{p_1}-\not\!{k_2}+m)
            \!\!\not{\epsilon^*_2}\right]u(p_1),\nonumber\\
{\cal M}_c&=&\frac{\kappa^2}{2(k_1\cdot k_2)}
           \left[(p_2\cdot\epsilon^*_2)(p_1\cdot\epsilon_1)
           -(p_1\cdot\epsilon^*_2)(p_2\cdot\epsilon_1)
           -(p_1\cdot k_1)(\epsilon_1\cdot \epsilon^*_2)
           \right]\nonumber\\
          &\times& \bar{u}(p_2)\left[\not\!{\epsilon_1}
           (k_1\cdot\epsilon^*_2)
           -\not\!{k_1}(\epsilon_1\cdot\epsilon^*_2)
           +\not\!{\epsilon^*_2}(k_2\cdot\epsilon_1)
           \right]u(p_1)\nonumber\\
          &-&\frac{\kappa^2}{2}[\epsilon_1\cdot\epsilon^*_2]
           \bar{u}(p_2)\left[\not\!{\epsilon_1}
           (p_1\cdot\epsilon^*_2)
           +\not\!{\epsilon^*_2}(p_2\cdot\epsilon_1)
           +\frac{\not\!{k_1}}{2}(\epsilon_1\cdot\epsilon^*_2)
           \right]u(p_1),\nonumber\\
{\cal M}_d&=&\frac{\kappa^2}{2}[\epsilon_1\cdot\epsilon^*_2]
           \bar{u}(p_2)\left[\frac{1}{2}\not\!{\epsilon_1}
           (\not\!{p_1}-\not\!{k_2}+m)\!\not\!{\epsilon^*_2}
           +\not\!{\epsilon_1}(p_2\cdot\epsilon^*_2)
           +\not\!{\epsilon^*_2}(p_1\cdot\epsilon_1)
           -\frac{\not\!{k_1}}{2}(\epsilon_1\cdot\epsilon^*_2)
           \right]u(p_1),
\end{eqnarray}
where $\epsilon^\mu_1\epsilon^\nu_1$ and $k^\mu_1$ are the
initial polarization and momentum of the graviton,
$\epsilon^{*\mu}_2\epsilon^{*\nu}_2$ and $k^\mu_2$ are the
final polarization and momentum of the graviton,
$p^\mu_1$ and $p^\mu_2$ are the initial and final momenta
of the massive fermion.
The transition amplitude ${\cal M}_{gg}$ is invariant under
the gravitational gauge transformation
\begin{eqnarray}
\epsilon^\mu_i\rightarrow \epsilon^\mu_i+\Lambda_i k^\mu_i,\ \
i=1,2.
\end{eqnarray}
where $\Lambda_i$ $(i=1,2)$ are arbitrary scalar functions.
The combination of two complicated ${\cal M}_c$ and ${\cal M}_d$
leads to a simple expression as
\begin{eqnarray}
{\cal M}_e&=&\frac{\kappa^2}{2(k_1\cdot k_2)}
           \left[(p_2\cdot\epsilon^*_2)(p_1\cdot\epsilon_1)
           -(p_1\cdot\epsilon^*_2)(p_2\cdot\epsilon_1)
           -(p_1\cdot k_2)(\epsilon_1\cdot \epsilon^*_2)
           \right]\nonumber\\
          &\times& \bar{u}(p_2)\left[\not\!{\epsilon_1}
           (k_1\cdot\epsilon^*_2)
           -\not\!{k_1}(\epsilon_1\cdot\epsilon^*_2)
           +\not\!{\epsilon^*_2}(k_2\cdot\epsilon_1)
           \right]u(p_1)\nonumber\\
          &+&\frac{\kappa^2}{4}
           [\epsilon_1\cdot\epsilon^*_2]
           \bar{u}(p_2)\left[\not\!{\epsilon_1}
           (\not\!{p_1}-\not\!{k_2}+m)\!\!\not\!{\epsilon^*_2}
           \right]u(p_1).
\end{eqnarray}
On the other hand, all three amplitude parts (${\cal M}_a$,
${\cal M}_b$ and ${\cal M}_e$) are not independent.
Transversality of graviton polarization yields
the relation,
\begin{eqnarray}
&&\bar{u}(p_2)\left[\not\!{\epsilon}_1(k_1\cdot \epsilon^*_2)
 -\not\!{k}_1(\epsilon_1\cdot\epsilon^*_2)+\not\!{\epsilon}^*_2
 (k_2\cdot \epsilon_1)\right]u(p_1)\nonumber\\
&&\hskip 4cm =\frac{1}{2}\bar{u}(p_2)
  \left[\not\!{\epsilon}^*_2(\not\!{p}_1
  +\not\!{k}_1+m)\!\not\!{\epsilon}_1
  -\not\!{\epsilon}_1(\not\!{p}_1-\not\!{k}_2+m)
  \!\not\!{\epsilon}^*_2\right]
  u(p_1),\label{identity}
\end{eqnarray}
and Eq.\ (\ref{identity}) renders the transition amplitude
${\cal M}_{gg}$ completely factorized,
\begin{eqnarray}
{\cal M}_{gg}&=&-\frac{\kappa^2}{2}F
     \left[(\epsilon_1\cdot\epsilon^*_2)
     -\frac{(p_1\cdot\epsilon_1)
      (p_2\cdot\epsilon^*_2)}{(p_1\cdot k_1)}
     +\frac{(p_2\cdot\epsilon_1)
     (p_1\cdot\epsilon^*_2)}{(p_1\cdot k_2)}
     \right]\nonumber\\
     &\times&\hskip 0.2cm\bar{u}(p_2)
      \left[\frac{\not\!{\epsilon}^*_2
     (\not\!{p}_1+\not\!{k}_1+m)
      \!\not\!{\epsilon}_1}{2(p_1\cdot k_1)}
     -\frac{\not\!{\epsilon}_1(\not\!{p}_1-\not\!{k}_2+m)
      \!\not\!{\epsilon}^*_2}{2(p_1\cdot k_2)}\right]u(p_1),
      \label{factor}
\end{eqnarray}
where the overall kinematical factor $F$ is
\begin{eqnarray}
F=\frac{(p_1\cdot k_1)(p_1\cdot k_2)}{(k_1\cdot k_2)},
\end{eqnarray}
and two bracket factors are nothing but the transition amplitudes
of scalar and fermion QED Compton scattering processes,
respectively.
The conclusion is that the gravitational Compton scattering
amplitude is the direct folding of scalar and fermion QED Compton
scattering amplitudes with an overall kinematical factor $F$.

The factorization property is not only confined to the linearized
gravity but it is quite general. As a matter of fact,
any non-Abelian theory with a semi-simple gauge group has
similar factorization properties \cite{Goebel}. For instance,
the transition amplitude of a gluon scattering with a colored
massive quark also is completely factorized into one part dependent
only on the non-Abelian gauge group structure as well as kinematics and
the other part which is nothing but the fermion QED Compton
scattering amplitude.
The group-dependent part takes a simple expression in terms of
an anticommutator and a commutator of group generators $T^a$
and $T^b$ as
\begin{eqnarray}
\{T^a,T^b\}_{ij}
-\frac{p_1\cdot (k_1+k_2)}{(k_1\cdot k_2)}[T^a,T^b]_{ij},
\label{QCDF}
\end{eqnarray}
where $i$ and $j$ are color indices for the initial and final
fermions, $a$ and $b$ are gluon indices for the initial and
final gluons, respectively.
Moreover, the linearized gravity coupled with QED has the
amplitude ${\cal M}_{\gamma g}$ of the graviton photoproduction
$\gamma e\rightarrow g e$ factorized \cite{Choi}
\begin{eqnarray}
{\cal M}_{\gamma g}=\frac{\kappa}{4e}
            [n\cdot \epsilon^*_g]
            {\cal M}_{\gamma\gamma},
\end{eqnarray}
where $e$ and ${\cal M}_{\gamma\gamma}$ are the electron charge
and the QED Compton scattering amplitude, respectively, and
\begin{eqnarray}
n^\mu=\left[(p_1+p_2)^\mu
     -\frac{p_1\cdot (k_1+k_2)}{(k_1\cdot k_2)}(k_1+k_2)^\mu\right].
     \label{LQED}
\end{eqnarray}

In brief, each amplitude of a gauge boson scattering
with a massive fermion is completely factorized into two parts
in any theory with unbroken gauge symmetries, where one
spin-dependent part is of the same form as the well-known QED
Compton scattering amplitude and the other is independent
of the spin of matter fields.
The independence of matter spin is still preserved in the linearized
gravity. But, it is one unique characteristic of the gravitational
Compton scattering that both amplitude factors of the process
depend on graviton polarization, whereas one amplitude factor of
any other process does not depend on gauge boson polarization.

Let us investigate the graviton polarization dependence of
scalar QED Compton amplitude in Eq.\ (\ref{factor}) in more detail.
Masslessness requires a graviton to have only two helicity states.
One useful polarization basis \cite{Voronov,Landau} consists of
two four-vectors
\begin{eqnarray}
n^\mu_1=\frac{1}{2}Nn^\mu,\ \
n^\mu_2=N\frac{\varepsilon^{\mu\nu\alpha\beta}p_{1\nu}
           k_{2\alpha}k_{1\beta}}{(k_1\cdot k_2)},
\end{eqnarray}
with the normalization factor $N=1/\sqrt{2F-m^2}$.
The vectors $n_1$ and $n_2$ satisfy the following relations:
\begin{eqnarray}
n_i\cdot n_j&=&-\delta_{ij},\ \
p_i\cdot n_2=k_j\cdot n_i=0,\nonumber\\
p_i\cdot n_1&=&\left(\frac{su-m^2}{t}\right)^{1/2},\ \ (i,j=1,2),
\end{eqnarray}
where $s$, $t$, and $u$ are Mandelstam variables.
The basis enables us to simultaneously construct the polarization
vectors $\epsilon_1(\lambda)$ and $\epsilon^*_2(\lambda)$
($\lambda=\pm 1$) for the initial and final gravitons as
\begin{eqnarray}
\epsilon_1(\lambda)=\epsilon^*_2(\lambda)=
\frac{1}{\sqrt{2}}(n_1+i\lambda n_2).
\end{eqnarray}
Any basis change gives rise to only an overall phase and
so it gives no effect on any physical observable. In the amplitude,
only the direct inner product $[\epsilon_1\cdot\epsilon^{*}_2]$
is dependent on graviton polarization:
\begin{eqnarray}
\epsilon_i(\lambda)\cdot\epsilon^*_2(\lambda^\prime)
 =-\delta_{\lambda,-\lambda^\prime}.\label{dependence}
\end{eqnarray}
Conventional gauge theories do not show this polarization dependence
and hence the polarization difference could be utilized for
discriminating between graviton interactions and other gauge-boson
interactions.

Both the dependence (\ref{dependence}) on graviton polarization
and the comparison of (\ref{LQED}) with (\ref{QCDF}) imply that
a four-momentum serves as a gravitational charge, while graviton
helicities serve as gauge-group indices in the linearized gravity.
On the other hand, the overall kinematical factor $F$ reads
in the center of mass frame
\begin{eqnarray}
F=\frac{1}{2}\left[s \frac{1+\cos\theta}{1-\cos\theta}+m^2\right],
\ \ \theta={\rm angle}\ \ {\rm between\ \ gravitons},
\end{eqnarray}
and its linear $s$-dependence causes unitarity violation at high
energies.

In conclusion, the factorization  is a common property of every
field theory with unbroken local symmetries.
The factorization maintenance of Einstein gravity
strengthens much more that the theory should be the very
low-energy limit of the ultimate guantum gravitational
theory \cite{Weinberg,Boul}.
While it might be at present beyond reach to take a great
conceptual leap toward the ultimate gravitational theory,
it looks clear that some crucial modifications must be considered
in factors, which are independent of the spin of matter fields.

\section*{Acknowledgments}

The work is supported in part by the Korea Science and Engineering
Foundation through the SRC program and in part by the SNU Daewoo
Research Fund.

\begin{figure}
\caption{Feynman diagrams for the gravitational Compton scattering process.
        The curly line is for a graviton and the solid line for
        a massive fermion.}
\label{Feynman Diagrams}
\end{figure}
\end{document}